\documentclass[reprint, prmaterials, amsfonts, amssymb, amsmath, aps, floatfix, article]{revtex4-2}
\usepackage[pdfusetitle]{hyperref}
\usepackage{float}
\usepackage{graphicx}
\hypersetup{
    colorlinks=true,
    linkcolor=blue,
    filecolor=blue,
    citecolor=blue,
    urlcolor=blue,
    }
\usepackage{longtable}

\usepackage{xpunctuate}
\DeclareRobustCommand\etal{\xperiodafter{\emph{et al}}}

\begin{document}

\title{Impact of hole-doping on the thermoelectric properties of pyrite FeS$_2$}
\author{Anustup Mukherjee}
    \email[Contact: ]{anustup.mukherjee@polytechnique.edu}
    \affiliation{CPHT, CNRS, \'Ecole Polytechnique, Institut Polytechnique de Paris, 91128 Palaiseau, France} 
\author{Alaska Subedi}
    \affiliation{CPHT, CNRS, \'Ecole Polytechnique, Institut Polytechnique de Paris, 91128 Palaiseau, France}

\date{\today} 

\begin{abstract}

We present a comprehensive first-principles analysis of the thermoelectric transport properties of 
hole-doped pyrite FeS$_2$ that includes electron-phonon interactions. This work was motivated by the
observed variations in the magnitude of thermopower reported in previous experimental and 
theoretical studies of hole-doped FeS$_2$ systems. Our calculations reveal that hole-doped
FeS$_2$ exhibits large positive room-temperature thermopower across all doping levels, 
with a room-temperature thermopower of 608 $\mu$V/K at a low hole-doping concentration of 10$^{19}$ cm$^{-3}$. 
This promising thermopower finding prompted a comprehensive investigation of other 
key thermoelectric parameters governing the thermoelectric figure of merit $ZT$. 
The calculated electrical conductivity is modest and remains below 10$^5$ S/m at 
room-temperature for all doping levels, limiting the achievable power factor. Furthermore, 
the thermal conductivity is found to be phonon driven, with a high room-temperature lattice thermal 
conductivity of 40.5 W/mK. Consequently, the calculated $ZT$ remains below 0.1, 
suggesting that hole-doped FeS$_2$ may not a viable candidate for effective thermoelectric 
applications despite its promising thermopower.

\end{abstract}


\maketitle

\section{Introduction}

The need of the hour is to use sustainable materials for energy applications, and FeS$_2$ in 
its pyrite form is a promising candidate. It has an indirect band gap of 0.95 eV \cite{Schlegel1976,Ennaoui1985}, 
which has motivated different scientific groups to investigate the electronic and transport properties 
of self- and impurity-doped FeS$_2$ for photovoltaic \cite{Ennaoui1993, Tomm1995, Eyert1998}, 
battery \cite{Strauss2000}, and thermoelectric \cite{Kato1997, Uhlig2014} applications. Various dopants such as 3$d$ transition metals 
\cite{Thomas1999,Uhlig2014,Diaz-Chao2008,Lehner2006,Mukherjee2024}, halogens \cite{Hu2012} and 
pnictogens \cite{Hu2012,Lehner2006,Lei2021} have been used to optimize the 
transport properties of pyrite FeS$_2$ by controlling the carrier concentration and nature of the charge carrier.

Over the years, the thermoelectric properties of electron-doped FeS$_2$ have been extensively 
studied through both experimental and theoretical investigations \cite{Uhlig2014,Clamagirand2016,Diaz-Chao2008,Gudelli2013,Mukherjee2024}. 
However, electron doping has proven to be an ineffective strategy for enhancing thermoelectric performance. 
Experimental data consistently show relatively low thermopower ($S$) in electron-doped FeS$_2$. The 
majority experimental studies on Co-doped and low-mobility natural n-type samples exhibit room-temperature $S$ 
below $-100$ $\mu$V/K \cite{Thomas1999,Uhlig2014,Diaz-Chao2008,Clamagirand2016,Willeke1992}, although two studies 
on natural n-type samples have reported absolute room-temperature $S$ values 
in the range of 300--400 $\mu$V/K \cite{Kato1997,Karguppikar1988}.
The relatively low value of $S$ in electron-doped FeS$_2$ is primarily attributed to the presence of a highly dispersive band at the bottom of the
conduction band manifold in the electronic structure of FeS$_2$, resulting in low density 
of states (DOS) and consequently reduced $S$.
In contrast, 
the electronic structure of FeS$_2$ reveals 
multiple flat 
bands near the valence band edge, contributing to a high DOS in this region \cite{Eyert1998,Zhao1993,Harran2017,Folkerts1987,Mukherjee2023}. These heavy  bands 
and increased DOS may result in larger $S$ and, consequently, a higher thermoelectric figure 
of merit because $ZT$ $\propto$ $S^2$. Therefore, it is reasonable to anticipate 
that hole-doped pyrite FeS$_2$ could exhibit promising thermoelectric properties.
Despite this potential, there is a relative scarcity of experimental and theoretical studies 
exploring the electronic and thermoelectric properties of hole-doped FeS$_2$.


A density functional theory (DFT) study by Gudelli \etal reported a large thermopower of approximately 750 $\mu$V/K 
at a hole concentration of $10^{19}$ cm$^{-3}$ for temperatures up to 800 K, with a reduction to 375 $\mu$V/K
at 900 K due to bipolar conduction \cite{Gudelli2013} . Another DFT study by Harran \etal  predicted a large $S$ 
of up to 567 $\mu$V/K  for hole concentrations of 5$\times$10$^{19}$ to 
1$\times$10$^{20}$ cm$^{-3}$ below 500 K, and a $ZT$ of $\sim$ 0.212 at room-temperature 
for a doping level of 10$^{20}$ cm$^{-3}$ \cite{Harran2017}. 
Even in undoped samples, experimental studies have observed relatively large positive thermopower,
possibly due to defects and impurities present on the surfaces of samples. 
Karguppikar and Vedeshwar  found natural p-type samples exhibiting $S$ between 430$-$660 $\mu$V/K \cite{Karguppikar1988}. 
Others have reported more modest values of $S$.  Uhlig \etal  measured 
$S$ of 128 $\mu$V/K at room temperature in nanoscale pyrite FeS$_2$ \cite{Uhlig2014}, while 
Rehman \etal  observed $S$ of 119 $\mu$V/K at 373 K 
in lab-grown pyrite FeS$_2$ nanoparticles \cite{Rehman2020}. Harada \etal and Zu\~niga-Puelles \etal  
reported positive $S$ below 300 K, followed by a change of sign at higher temperatures, leading to 
large negative values \cite{Harada1998,Zuniga2019}. These significant variations in measured $S$ underscore 
the need for a more detailed investigation into the thermoelectric transport properties of hole-doped pyrite FeS$_2$.

In this paper, we examine the thermoelectric properties of hole-doped pyrite FeS$_2$ using 
first-principles calculations that incorporate scattering processes due to electron-phonon interactions. We find
positive $S$ below 400 K across all examined hole concentrations, with a maximum room-temperature 
$S$ of approximately 608 $\mu$V/K at the lowest studied hole doping level (10$^{19}$ cm$^{-3}$). $S$ changes 
sign due to bipolar conduction effects in the investigated temperature range, except 
for the highest doping level. Electrical conductivity ($\sigma$) exhibits a non-monotonic trend, 
initially decreasing with temperature before increasing at higher
temperatures, 
also influenced by bipolar conduction. The room-temperature $\sigma$ remains 
around 10$^{5}$ S/m irrespective 
of the doping level, with a value of $\sim$ 470 S/m at hole doping 
level of $10^{19}$ cm$^{-3}$. 
Our calculations taking account the electron-phonon and three-phonon interactions indicate that the thermal conductivity is predominantly phonon-driven, with the lattice 
contribution ($\kappa_{ph}$) exceeding 40 W/mK at room-temperature, while the electronic thermal 
conductivity ($\kappa_e$) remains below 0.15 W/mK across all doping levels at 300 K. We obtain a 
reasonably good power factor (PF) at 100 K, but relatively low PF between 300$-$700 K for 
low doping levels. The thermoelectric figure of merit remains below 0.1 across all temperatures 
and doping levels, primarily due to low electrical conductivity and high lattice thermal 
conductivity, as $ZT$ is proportional to $\sigma/(\kappa_e + \kappa_{ph})$.
Despite the promisingly high value of  
thermopower, the overall low $ZT$ suggests that hole-doped pyrite FeS$_2$ 
is unlikely to exhibit 
high thermoelectric efficiency. While our calculations provide valuable insights into the 
thermoelectric properties of hole-doped pyrite FeS$_2$, experimental investigations are 
crucial to validate these theoretical predictions and explore potential avenues for improving 
its thermoelectric performance.

\section{Computational Details}

We performed density functional theory calculations 
using the 
{\sc quantum espresso} (QE) package \cite{Giannozzi2009,Giannozzi2017,Giannozzi2020}. The exchange-correlation part of the Hamiltonian 
was approximated by the GGA scheme of Perdew, Burke, and Ernzerhof \cite{Perdew1996}. 
We employed the ultrasoft pseudopotentials with core corrections  
obtained from {\sc pslibrary} (v.1.0.0) \cite{Dal-Corso2014}. 
The valence shell  electronic configurations of the pseudopotentials were $3d^{6}4s^{2}$ (Fe) and $3s^{2}3p^{4}$ (S).
A converged kinetic energy and charge density cut-off 
of 60 and 600 Ry were taken for the self-consistent cycles, respectively. A $\Gamma$-centered 
$k$-point mesh of $12 \times 12 \times 12$ 
was used for the self-consistent calculation.  
The energy convergence criterion was set to $10^{-14}$ Ry.



To calculate the phonon dispersion and the second-order interatomic force constants, 
we employed density functional perturbation theory (DFPT) implemented in the QE code. 
We used 
a $\Gamma$-centered $q$-point mesh of $4 \times 4 \times 4$ to obtain the dynamical matrices. This mesh 
is consistent with the $k$-point mesh used for non-self-consistent calculation, as required by the 
{\sc perturbo} code \cite{Zhou2021} to calculate the transport properties. {\sc perturbo} solves the 
semiclassical Boltzmann transport equation (BTE) taking electron-phonon (e-ph) interactions into account. 
Within the Boltzmann transport theory, the BTE for a periodic system is formulated as:
\begin{equation}
\begin{aligned}
    \label{BTE}
     \frac{\partial f_{nk}(r,t)}{\partial t} & = -[\nabla_r f_{nk}(r,t).v_{nk} + {\hbar}^{-1} \nabla_k f_{nk}(r,t).F] \\
     & + I[f_{nk}],
\end{aligned}
\end{equation}
where $f_{nk}(r,t)$, $t$, $k$, $r$, $n$, $v_{nk}$, $\hbar$, $F$ and $I[f_{nk}]$ are the electron occupations, 
time, crystal momentum, spatial coordinate, band index, band velocities, reduced Planck's constant, external fields,
and collision integral, respectively. 
The collision integral is given by:
\begin{equation}
\begin{aligned}
    \label{CollisionIntegral}
     I[f_{nk}] &= -\frac{2\pi}{\hbar}\frac{1}{N_q}\sum\limits_{mqv}{\vert g_{mqv}(k,q) \vert}^{2} \\
     & \times [\delta (\epsilon_{nk} - \hbar\omega_{vq} - \epsilon_{mk+q}) \times F_{em}] \\
     & \times [\delta (\epsilon_{nk} + \hbar\omega_{vq} - \epsilon_{mk+q}) \times F_{abs}],
\end{aligned}
\end{equation}
where $N_q$, $q$, $\vert g_{mqv}(k,q) \vert$, and $v$ represent the number of $q$-points used for the summation, 
phonon wavevector, e-ph matrix elements, and mode index, respectively. The electron quasiparticles 
and phonon energies 
are given as $\epsilon_{nk}$ and $\hbar\omega_{vq}$. $F_{em}$ and $F_{abs}$ represent the 
phonon emission and absorption terms, respectively.

\begin{figure*}[!ht]
    \includegraphics[width=\textwidth,height=0.37\textwidth]{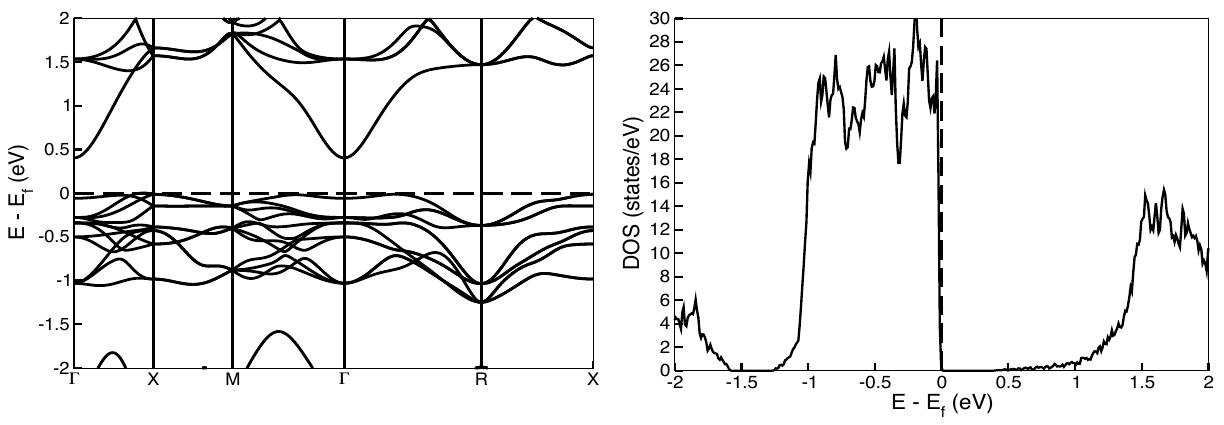}
    \caption{(Left) Electronic band structure and (right) density of states of stoichiometric FeS$_2$. The Fermi energy is shifted to the valence band maximum.}
    \label{fig:FeS2 band structure dos}
\end{figure*}

{\sc perturbo} reads the results of DFT and DFPT calculations performed on coarse $k$- and 
$q$-point grids and then expresses the 
electron energies, phonon frequencies and e-ph matrix elements in the Wannier basis. 
Transport properties are calculated by interpolating these quantities
on fine $k$- and $q$-point grids and using them in the solution of the BTE. We used an interpolation factor of 80 
and performed an uniform sampling of the first Brillouin zone using 500,000 random $q$-points to 
obtain the relaxation time $\tau_{nk}$. We then applied the relaxation time approximation (RTA) 
to calculate $S$, $\sigma$ and $\kappa_e$. The conductivity and Seebeck tensors are given by:
\begin{align}
    \label{Conductivity}
     \sigma_{\alpha\beta} &= e^{2}\int dE(-\partial f^0/\partial E) \sum_{\alpha\beta}(E) \\
     \label{Seebeck}
     [\sigma S]_{\alpha\beta} &= \frac{e}{T}\int dE(-\partial f^0/\partial E)(E-\mu) \sum_{\alpha\beta}(E).
\end{align}
Here $S$ is the Seebeck coefficient and $\sum_{\alpha\beta}(E)$ is the transport distribution function (TDF), which is defined as:
\begin{align}
     \label{TDF}
      \sum_{\alpha\beta}(E) &= \frac{s}{N_k\Omega}\sum_{nk}{v_{nk}^{\alpha}}{F_{nk}^{\beta}}\delta(E-\epsilon_{nk}),
\end{align}
where $e$ is the electronic charge, $\alpha$ and $\beta$ are the Cartesian directions, $f^0$ represents the equilibrium Fermi-Dirac distribution, $\mu$ is the chemical potential, $T$ is the temperature, $s$ represents the spin degeneracy, $\Omega$ is unit cell volume, and $F_{nk} = \tau_{nk}v_{nk}$ in the RTA approach. 

The third-order force constants used in the calculation of the lattice contribution to thermal conductivity 
was obtained using the {\sc thirorder.py} code \cite{Li2012}. We used a $3 \times 3 \times 3$ supercell to generate an irreducible set of 
atomic displacements up to the third nearest neighbour.  Self-consistent DFT runs were performed on 
the generated set to obtain the atomic forces, which were then used to calculate the third-order force constants. 
The lattice thermal conductivity  
was then calculated by solving the Boltzmann transport equation for phonons within RTA
using the ShengBTE package \cite{Li2012,Li2014}. $\kappa_{ph}$ within BTE is formulated as:
\begin{align}
     \label{kappa_l}
      {{\kappa}_l}^{\alpha\beta} &= \frac{1}{k_{b}T^{2}\Omega N}\sum_{\lambda}f_{0}(f_{0}+1)(\hbar \omega_{\lambda})^{2}{v_\lambda^\alpha}{F_\lambda^\beta},
\end{align}
where $k_b$, $N$, $f_0$, $\omega_{\lambda}$, $\lambda$, and 
${v_\lambda^\alpha}$
 are Boltzmann's constant, number of $q$-points, Bose-Einstein distribution 
function, angular frequency, phonon mode, and group velocity, respectively. 
${F_\lambda^\beta} = {\tau_\lambda^0} v_\lambda$ is the linearized BTE within RTA and ${\tau_\lambda^0}$
is the relaxation time of the phonon mode $\lambda$.

\section{Structural Details}

Pyrite FeS$_2$ adopts a cubic crystal structure, with a space group of $Pa\overline{3}$, 
with four formula units per unit cell. The Fe and S atoms occupy distinct crystallographic sites 
and their relative positions are governed by a single structural parameter $u$.
For our calculations, 
we used an experimentally determined lattice constant of $a$ = 5.407 \AA, as reported in previous studies 
\cite{Umemoto2006,Wyckoff1963}. The atomic positions were then obtained by minimizing the atomic forces 
to less than 10$^{-6}$ Ry/Bohr. The optimized value for $u$, which governs all
the bond distances, was found to be 0.3827.

\section{Results and Discussions}

\subsection{Electronic Structure}

Numerous density functional theory studies have previously explored the electronic structure 
of stoichiometric FeS$_2$ \cite{Schlegel1976, Zhao1993, Eyert1998, Umemoto2006, Feng2018, Mukherjee2023}. 
These calculations consistently reveal 
multiple heavy bands near the valence band edge that are composed of  
nominally Fe $t_{2g}$ orbitals, as well as a highly dispersive band at the bottom of the conduction band manifold featuring
a mix of Fe $e_g$ and S $p$ character. Consequently, high electronic density of states 
are observed near the valence band edge, while the conduction band region shows a relatively lower 
DOS. 
In Fig.~\ref{fig:FeS2 band structure dos}, we reproduce the band structure and 
DOS of FeS$_2$ in the left and right columns, respectively. 
Our calculations yield an indirect band gap of 0.40 eV using the PBE functional, which is in reasonable agreement with the 
0.52 eV band gap reported by Harran \etal using the HSE functional \cite{Harran2017}. However, 
our computed band gap is lower than the 0.95 eV value reported in previous experimental 
studies \cite{Schlegel1976,Ennaoui1985}.

\subsection{Transport Properties}

\begin{figure}[htb]
    \includegraphics[width=\columnwidth]{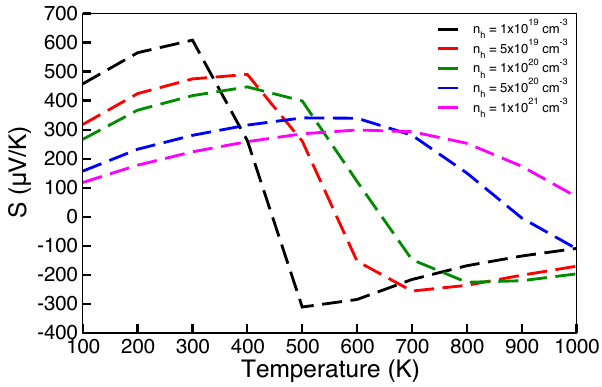}
    \caption{Seebeck coefficient $S$ of hole-doped FeS$_2$ in the concentration range 10$^{19}$--10$^{21}$ cm$^{-3}$.}
    \label{fig:seebeck FeS2 hole doped}
\end{figure}

We first look at the temperature-dependent thermopower $S$ of hole-doped FeS$_2$ across various doping levels, 
as shown in Fig.~\ref{fig:seebeck FeS2 hole doped}. Our calculations indicate a consistent behaviour 
in $S$ for hole doping concentrations ($n_h$) at and below 1$\times$10$^{20}$ cm$^{-3}$. Initially, $S$ increases 
with increasing temperature. However, at higher temperatures, it begins to decrease, eventually 
reaching large negative values due to occupation of dispersive conduction band across the gap. 
With further increase in temperature, $S$ becomes less negative within this doping range. 
For $n_h$ of 5$\times$10$^{20}$ cm$^{-3}$, the Fermi level now lies deeper.  Consequently, $S$ 
becomes negative above only above 800 K. A similar behaviour in $S$ is also reported in several experiments \cite{Harada1998,Zuniga2019}.
In contrast, for the highest $n_h$ considered in this study (1$\times$10$^{21}$ cm$^{-3}$),
$S$ 
remains positive within the investigated temperature range. However, we predict $S$ to change its sign for this doping level
at higher temperatures 
when bipolar conduction becomes dominant.
The room-temperature $S$ values for the studied hole doping levels are reported in Table \ref{Room temperature Seebeck conductivity table}. 
The maximum room-temperature $S$ of 608 $\mu$V/K occurs, as expected, for the lowest doping concentration 
of 1$\times$10$^{19}$ cm$^{-3}$. With further increase in hole doping, the room-temperature $S$ 
decreases gradually but remains positive. This behaviour can be rationalized by observing the 
electronic structure of FeS$_2$. As more holes are introduced in the system, the Fermi energy 
shifts deeper in the valence band manifold where the bands are more dispersive. This increases particle-hole symmetry, thereby reducing $S$.
Since the calculated values of $S$ are promising, with room-temperature $S$ 
exceeding 200 $\mu$V/K across all studied $n_h$ values, we extend our analysis to the additional 
transport parameters that govern $ZT$.

\begin{table}[t]
\caption{Room temperature thermopower $S$ and electrical conductivity $\sigma$ of hole-doped FeS$_2$ at doping concentrations $n_h$ ranging from 10$^{19}-$10$^{21}$ cm$^{-3}$.
}
\label{Room temperature Seebeck conductivity table}
\begin{ruledtabular}
\begin{tabular}{c c c c c}
$n$ (cm$^{-3}$) & $S$ ($\mu$V/K) & $\sigma \times 10^{2}$ (S/m) \\  \hline
1 $\times$ 10$^{19}$ & $608.12$ & $4.72$ \\
5 $\times$ 10$^{19}$ & $474.72$ & $23.71$ \\
1 $\times$ 10$^{20}$ & $417.10$ & $46.27$ \\
5 $\times$ 10$^{20}$ & $280.53$ & $220.88$ \\
1 $\times$ 10$^{21}$ & $223.70$ & $416.16$ \\
\end{tabular}
\end{ruledtabular}
\end{table}

\begin{figure}[t]
    \includegraphics[width=\columnwidth]{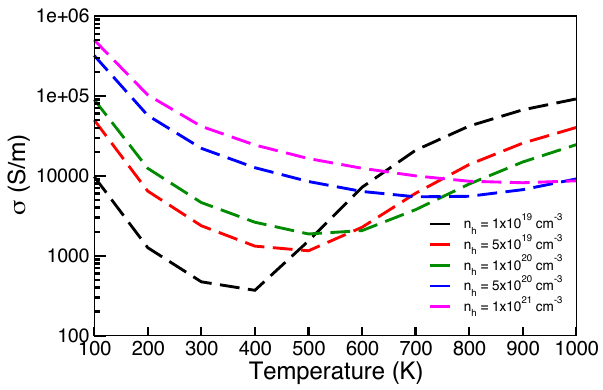}
    \caption{Electrical conductivity $\sigma$ of hole-doped FeS$_2$ in the concentration range 10$^{19}-$10$^{21}$ cm$^{-3}$.}
    \label{fig:conductivity FeS2 hole doped}
\end{figure}

The calculated room-temperature $\sigma$ for all doping levels are summarized in Table 
\ref{Room temperature Seebeck conductivity table}, and they show the expected increase 
in $\sigma$ as a function of $n_h$.  Fig.~\ref{fig:conductivity FeS2 hole doped} 
shows the temperature dependence of $\sigma$ for various hole doping levels.
We find that $\sigma$ shows a similar trend as a function of $n_h$ at and below 
1$\times$10$^{20}$ cm$^{-3}$.
Initially, $\sigma$ decreases with increasing temperature. 
However, further increase in temperature leads to upturn in $\sigma$ due to increased population of the dispersive 
conduction band. 
As the doping concentration is increased, the upturn in $\sigma$ occurs at higher values.   
For $n_h$ of 1$\times$10$^{19}$ cm$^{-3}$, the upturn happens at 400 K, where as $\sigma$ starts increasing 
only above 800 K for $n_h$  of 5$\times$10$^{20}$ cm$^{-3}$.
The increase in the population of the dispersive conduction band at high temperatures also 
leads to high-temperature $\sigma$ being larger for low doping levels that is observed in Fig.~\ref{fig:conductivity FeS2 hole doped}. 
Meanwhile, the Fermi level lies deeper for the highest considered $n_h$ of 1$\times$10$^{21}$ cm$^{-3}$. 
Consequently, the dispersive conduction band is less occupied for this case, and its 
$\sigma$ decreases continuously with temperature, with only a moderate increase observed above 900 K.

\begin{figure}[b]
    \includegraphics[width=\columnwidth]{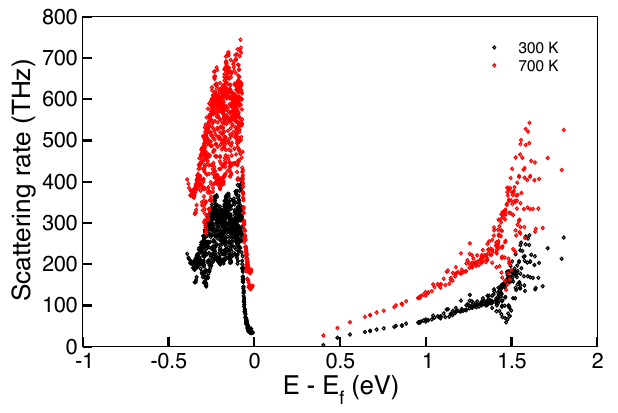}
    \caption{Electron-phonon scattering rates (the inverse of relaxation time) as a function of energy at 300 and 700 K for carrier concentration $n_h$ = 1$\times$10$^{19}$ cm$^{-3}$. The Fermi energy is shifted to the valence band maximum.}
    \label{fig:scattering rate el ph}
\end{figure}

We have so far discussed the temperature dependence of $S$ and $\sigma$ only in terms of the
relative difference in the dispersion of the heavy valence and light conduction bands near
Fermi level.  The relaxation time $\tau_{nk}$ of the carriers in these bands is another 
microscopic quantity that could play a role in the temperature dependence of these physical quantities. 
Fig.~\ref{fig:scattering rate el ph} shows the calculated el-ph scattering rates $( = 1/\tau_{nk})$
at 300 K and 700 K for $n_h$ of 1$\times$10$^{19}$ cm$^{-3}$. 
At 300 K, the e-ph scattering rates near the valence band edge are higher compared to those 
at the conduction band minimum. 
At 700 K, the scattering rates increase for both sets of bands, but the increase is much larger in the valence band 
manifold.  The scattering rates of the states near the valence band edge are now around an order of magnitude larger than
those near the conduction band edge.  This large asymmetry likely causes the large jump in $S$ to negative
values at high temperatures.  The relatively low values of scattering rate in the conduction band manifold also
contributes to the increase in $\sigma$ at high temperatures.

\begin{figure}[t]
    \includegraphics[width=\columnwidth]{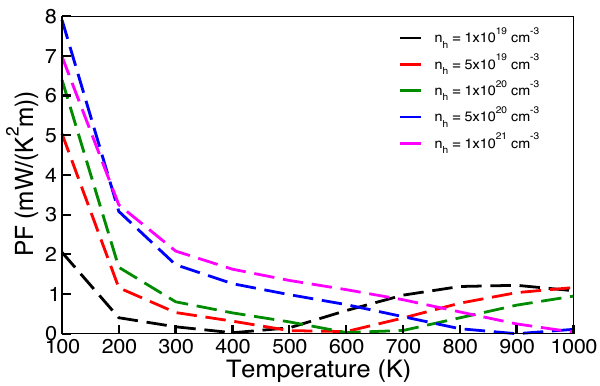}
    \caption{Power factor  of hole-doped FeS$_2$ in the concentration range 10$^{19}$--10$^{21}$ cm$^{-3}$.}
    \label{fig:Power Factor}
\end{figure}

\begin{figure}[b]
    \includegraphics[width=\columnwidth]{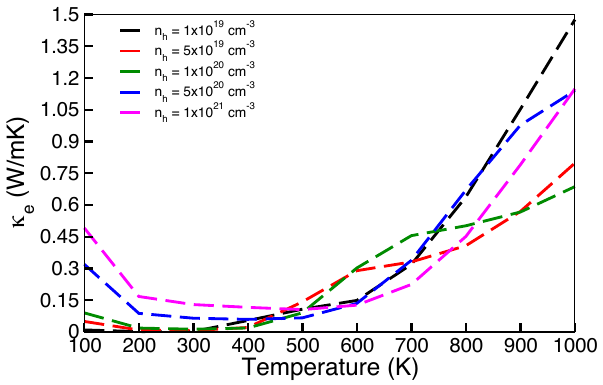}
    \caption{Electronic thermal conductivity $\kappa_e$ of hole-doped FeS$_2$ in the concentration range 10$^{19}$--10$^{21}$ cm$^{-3}$.}
    \label{fig:Electronic thermal conductivity}
\end{figure}

\begin{figure}[htb]
    \includegraphics[width=\columnwidth]{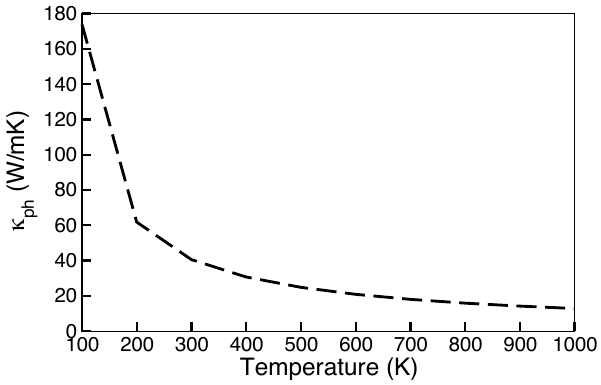}
    \caption{Lattice thermal conductivity $\kappa_{ph}$ of stoichiometric FeS$_2$ taking 
    into account three-phonon scattering interactions.}
    \label{fig:Lattice thermal conductivity}
\end{figure}

Following the observation of high values of calculated $S$, 
we investigated the power factor at the corresponding hole concentrations, which is  shown in Fig.~\ref{fig:Power Factor}
as a function of temperature. At 100 K, the PF is reasonably high, exceeding 2 mW/K$^{2}$m across 
all doping levels. This can be attributed to the moderately high conductivity at 100 K, given that PF 
is proportional to $\sigma$. However, for doping concentrations ranging between 10$^{19}$ 
and 10$^{20}$ cm$^{-3}$, the PF decreases and remains below 1 mW/K$^{2}$m within the temperature range of 
300 to 600 K. In this temperature range, $\sigma$ remains below $10^{5}$ S/m, and $S$ undergoes a sign change,
leading to low PF values. Only at the highest $n_h$ of 1$\times$10$^{21}$ cm$^{-3}$, the 
PF exceeds 1 mW/K$^{2}$m up to 600 K. Additionally, at 1000 K, we observe PF values close to 1 mW/K$^{2}$m for 
lower doping concentrations ($\leq$ 1$\times$10$^{20}$ cm$^{-3}$), which can be attributed to the 
increase in $\sigma$ and the large negative $S$ observed at this temperature.

Calculation of the thermoelectric figure of merit $ZT$ additionally requires the knowledge of thermal
conductivity.  
The electronic contribution to thermal conductivity $\kappa_e$ across all doping levels is shown in 
Fig.~\ref{fig:Electronic thermal conductivity}.  Our calculations reveal that $\kappa_e$ is relatively low, 
and remains below 1.5 W/mK across the entire temperature range studied.  Since the relatively small values of 
hole doping concentration considered here is should only modestly change the phonon dispersions, we calculated
the computationally demanding lattice contribution to thermal conductivity $\kappa_{ph}$ only for the undoped 
FeS$_2$, which is shown in Fig.~\ref{fig:Lattice thermal conductivity}.  In contrast to $\kappa_e$, 
$\kappa_{ph}$ is significantly higher, reaching a maximum of approximately 174 W/mK at 100 K.
The room-temperature $\kappa_{ph}$ is around 40.5 W/mK, whereas $\kappa_e$ reaches a maximum of 0.13 W/mK at the same temperature for a 
hole doping concentration of 1$\times$10$^{21}$ cm$^{-3}$.
This suggests that the thermal conductivity is predominantly phonon-driven at these temperatures. 
Our calculated room-temperature $\kappa_{ph}$ is in close agreement with
previous experimental finding 
of 47.8 $\pm$ 2.4 W/mK on single-crystal FeS$_2$ \cite{Popov2013}, and is lower than the 65.8 W/mK value 
reported in a previous theoretical study \cite{Jia2023}. 
However, a recent experimental
investigation reported a lattice thermal conductivity of approximately 22$–$24 W/mK at room-temperature 
in FeS$_2$ samples grown via chemical vapor transport (CVT) \cite{Ozden2023}. This discrepancy could be attributed to the 
presence of vacancies and surface impurities in the CVT-grown samples.

\begin{figure}[t]
    \includegraphics[width=\columnwidth]{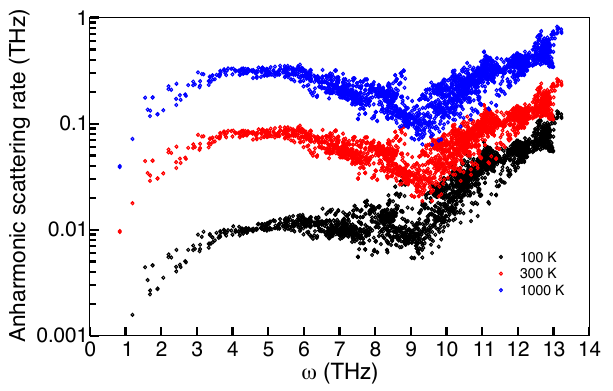}
    \caption{Anharmonic scattering rates of stoichiometric FeS$_2$ at temperatures of 100, 300 and 1000 K.}
    \label{fig:anharmonic scattering rate}
\end{figure}


\begin{figure}[b]
    \includegraphics[width=\columnwidth]{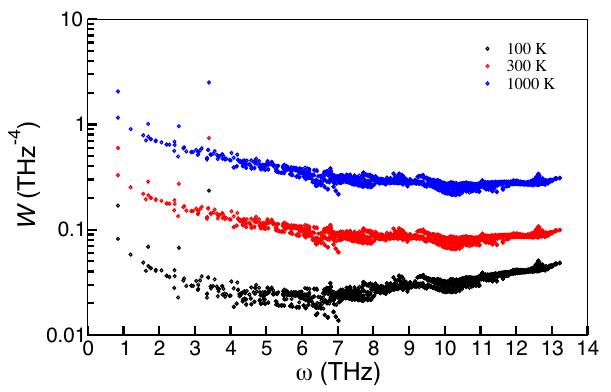}
    \caption{Weighted phase space $W$ of stoichiometric FeS$_2$ at 100, 300 and 1000 K.}
    \label{fig:weighted phase space}
\end{figure}

The high values of $\kappa_{ph}$ at low temperatures can be understood through an 
examination of anharmonic scattering rates. 
Fig.~\ref{fig:anharmonic scattering rate} illustrates the three-phonon contribution to anharmonic scattering rates 
obtained at 100, 300 and 1000 K. Our calculations show that anharmonic scattering rates 
increase with temperature across the entire frequency spectrum. However, this increase is particularly 
pronounced in the low to intermediate frequency region (below 8 THz). 
Since low frequency phonons are the primary carriers of heat, this substantial temperature-induced increase in scattering
rates results in a rapid decrease in $\kappa_{ph}$ as temperatures is increased.
Our calculated room-temperature
scattering rates predominantly lie in the range of 0.01$-$0.1 THz, comparable to those observed in skutterudites 
like CoSb$_3$ and IrSb$_3$ that also exhibit room-temperature $\kappa_{ph}$ 
values above 10 W/mK \cite{Wu2014}.  
In contrast, the skutterudite compound FeSb$_3$  shows significantly 
higher room-temperature scattering rates (0.1$-$10 THz), leading to an ultralow room-temperature 
$\kappa_{ph}$ of 1.14 W/mK \cite{Yuhao2016}. This comparison underscores the critical role of scattering rates 
in determining thermal transport properties. Our observation of the 
increased scattering rates 
at higher temperatures is further supported by the behaviour of the 
weighted phase space ($W$), 
as shown in Fig. \ref{fig:weighted phase space}. $W$ exhibits a similar 
temperature-dependent 
increase across the entire frequency range. This increase in $W$ indicates a greater availability 
of three-phonon scattering channels at elevated temperatures, which contributes to the 
observed reduction in $\kappa_{ph}$ at elevated temperatures.

\begin{figure}[htb]
    \includegraphics[width=\columnwidth]{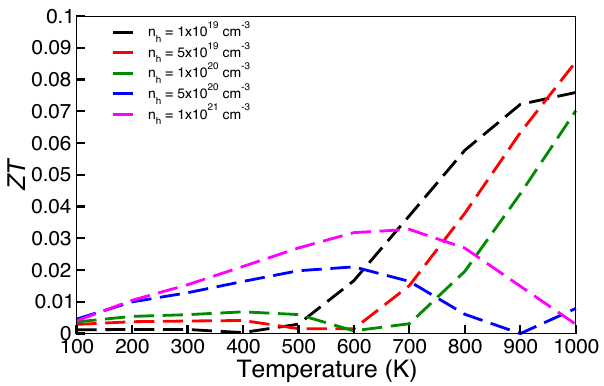}
    \caption{Thermoelectric figure of merit $ZT$ of hole-doped FeS$_2$ in the concentration range 10$^{19}-$10$^{21}$ cm$^{-3}$.}
    \label{fig:ZT}
\end{figure}

The thermoelectric figure of merit $ZT$ as a function of temperature obtained using the calculated
$S$, $\sigma$, and $\kappa$ is presented in Fig. \ref{fig:ZT}. $ZT$ is less than 0.01 at low temperatures
for all doping concentrations.  
Such low values can be attributed to the large lattice thermal conductivity obtained at low temperatures. 
As the temperature increases, $ZT$ shows a moderate increase up to 500 K for doping values between 
10$^{19}$ and 10$^{20}$ cm$^{-3}$, consistent with their low PF. At even higher temperatures, their $ZT$
increases due to a combination of increasing PF and decreasing $\kappa_{ph}$. 
For concentrations above 10$^{20}$ cm$^{-3}$, 
$ZT$ gradually increases with temperature, reaching a maximum between 600--700 K, and subsequently decreases
due to the low PF. However, for hole doping at 5$\times$10$^{20}$ cm$^{-3}$, a moderate increase 
in $ZT$ is observed above 900 K, which follows a similar increase in its PF. 
The room-temperature $ZT$ is small, with a maximum of 0.015 for the highest doping level of 1$\times$10$^{21}$ cm$^{-3}$. 
We obtain a highest $ZT$ of approximately 0.086
for hole concentration of 5$\times$10$^{19}$ cm$^{-3}$ at 1000 K. Our maximum calculated $ZT$ is 
lower than $ZT$ $\sim$ 0.212 at room temperature calculated by Harran \etal without taking into 
account electron-phonon scattering \cite{Harran2017}. 


\section{Conclusions}

In this study, we investigated the thermoelectric transport properties of hole-doped pyrite FeS$_2$, 
incorporating electron-phonon interactions across a wide doping range using first-principles calculations. 
This research was motivated by previous theoretical findings that revealed multiple heavy bands near 
the valence band edge in stoichiometric FeS$_2$, as well as subsequent reports of significant variations 
in thermopower in hole-doped FeS$_2$ from both experimental and theoretical perspectives.

Our investigation of hole-doped pyrite FeS$_2$ revealed complex thermoelectric behavior 
across various doping levels and temperatures. We observed a maximum positive $S$ of approximately 
608 $\mu$V/K at room temperature for the lowest doping level. Notably, bipolar conduction induced a sign change in $S$ at 
higher temperatures for all but the highest doping level. This bipolar conduction effect was also observed 
in the electrical conductivity, which, while moderately high at room temperature, 
remained below 10$^{5}$ S/m. The highest observed $\sigma$ was approximately 5$\times$10$^{5}$ S/m 
for the highest doping level at 100 K. Due to significant variations in $S$ and moderate $\sigma$ values, 
the power factor was relatively low above room temperature, exceeding 2 mW/K$^{2}$m only at the lowest 
temperatures across all doping levels. We find that thermal transport in the system is dominated by phonon 
contributions, with lattice thermal conductivity exhibiting remarkably high values 
below room temperature, including a room-temperature $\kappa_{ph}$ of 40.5 W/mK. 
This results from reduced anharmonic scattering rates and limited phase space availability at 
low temperatures. In contrast, the electronic contribution to the thermal conductivity remains 
low across all temperatures, consistently below 1.5 W/mK. The modest values of electrical conductivity and 
relatively high values of thermal conductivity lead to low $ZT$ of less than 0.1. This suggests that hole-doped 
pyrite FeS$_2$ is unlikely to be an effective material for thermoelectric applications even though it exhibits 
high calculated thermopower. So alternative doping strategies that improve electrical conductivity and degrade 
thermal conductivity should be explored to make this material viable for thermoelectric application.

\begin{acknowledgements}
We express our gratitude to Sylvie H\'ebert and David J. Singh for useful discussions. This work was 
supported by Agence Nationale de la Recherche under grant ANR-21-CE50-0033
and GENCI-TGCC under grant A0170913028. 
\end{acknowledgements}

\bibliography{references}

\end{document}